\documentclass[aps,twocolumn,showpacs,floatfix]{revtex4}
\usepackage{dcolumn}
\usepackage{graphicx}
\usepackage{amsfonts}
\usepackage{amsmath,bm,amssymb}
\usepackage{comment}
\usepackage[usenames]{color}

\def\vxc{V^{\rm xc}}
\def\ei{\varepsilon_i}

\def\ej{\varepsilon_j}
\def\bfk{{\bf k}}
\begin{document}

\title{Quasiparticle self-consistent $GW$ study of LaNiO$_3$ and
  LaNiO$_3$/LaAlO$_3$ superlattice}

\author{Myung Joon Han} \email{mj.han@kaist.ac.kr}
\affiliation{Department of Physics and KAIST Institute for the NanoCentury, Korea Advanced Institute of
  Science and Technology, Daejeon 305-701, Korea }
\author{Hiori Kino}
\affiliation{National Institute for Materials Science, Sengen 1-2-1, Tsukuba, Ibaraki 305-0047, Japan.}

\author{Takao Kotani}
\affiliation{Department of Applied Mathematics and Physics, Tottori University, Tottori 680-8552, Japan}
\date{\today }

\begin{abstract}
Using quasiparticle self-consistent $GW$ calculations we examined the
electronic structure of LaNiO$_3$ and the LaNiO$_3$/LaAlO$_3$
superlattice. The effects of electron correlation in Ni-$d$ bands were
reasonably well described without any {\it ad hoc} parameter and
without the ambiguity related to the double-counting and downfolding
issues. The effective mass is about 30\% enhanced compared to the GGA
result. One band feature, which is believed to be essential for the
cuprate-like superconductivity, is not realized and the central Fermi
surface pocket does not disappear. Our result is consistent with a
recent dynamical mean field calculation based on the $d$--$p$ model
and in contrast to the result from a $d$-band only model.
\end{abstract}


\maketitle

\section{Introduction}

Recently nickelate superlattices have attracted considerable
attention, especially due to the reported possibility of
high-temperature superconductivity. A recent dynamical mean field
theory (DMFT) calculation reported that the band structure of
LaNiO$_3$/LaAlO$_3$ (LNO/LAO)-type heterostructure can be cuprate-like
\cite{Hansmann} while another DMFT calculation based on a Hubbard
Hamiltonian explicitly containing oxygen $p$ bands made a different
prediction \cite{MJHan-DMFT}. Also, an intriguing metal-insulator
transition (MIT) accompanied by magnetic transition was observed
\cite{Boris-Science}.

From the theoretical point of view, it is a great challenge to
describe the correct ground state of LNO as well as a LNO/LAO
superlattice based on the first-principles method. Conventional LDA or
GGA overestimates the bandwidth and underestimates the effective mass
while the correct paramagnetic (PM) ground state is reproduced for LNO
\cite{Hamada}. This limitation is clearly manifested when it is
applied to LNO/LAO. While LNO/LAO becomes magnetic and insulating in
the thin LNO limit \cite{Boris-Science,Liu-PRB-RC,NNO-PRL,John-epl},
LDA/GGA still gives a PM and metallic ground state
\cite{MJHan-opol}. Since the LDA$+U$ type of approach strongly prefers
the magnetic ground state, it yields a local moment at the Ni site in
the LNO/LAO, consistent with experiments in the thin LNO limit
\cite{Boris-Science}, but the problem is that LDA$+U$ cannot describe
the correlated PM phase.  As a result, for the thick LNO limit of the
superlattice and for bulk LNO, LDA$+U$ predicts a ferro- (or
antiferro-) magnetic ground state, which is in sharp contrast with the
reality \cite{MJHan-LDAU}. Although DMFT is one of the ways to go
beyond LDA$+U$ especially for the correlated PM phase, there is a
discrepancy between two DMFT results for the Fermi surface topology of
the LNO/LAO superlattice as mentioned above
\cite{Hansmann,MJHan-DMFT}. This demonstrates the limitation of the
current DMFT scheme as an ideal first-principles electronic structure
calculation method. The ambiguities inevitably arise from the
downfolding, projection, $U$ and $J$ parameters, and double-counting
terms. From this perspective, the nickelate superlattice is an
archetypal example that challenges the predictive power of current
first-principles methods to simulate correlated electron systems.

Here we take an alternative approach, namely, the quasiparticle
self-consistent $GW$ (QSGW) method. Without any {\it ad hoc}
parameter, QSGW gives a reasonable band structure in terms of the
bandwidth and the effective mass both for LNO and the
superlattice. Importantly, our calculation shows that the band
structure of LNO/LAO does not become cuprate-like, which supports the
conclusion of the DMFT result based on the $d$--$p$ model.

\section{Computation Method}
\subsection{Quasiparticle self-consistent $GW$ method}

The QSGW was originally introduced by Faleev, van Schilfgaarde and
Kotani \cite{QSGW-PRL2004}, and has now become a widely used standard
method with which one can calculate $H_0$ (non-interacting Hamiltonian
describing quasiparticles (QPs) or band structures) and $W$
(dynamically-screened Coulomb interactions between the QPs within the
random phase approximation (RPA)) in a self-consistent manner.  Note
that QSGW fully takes into account the non-locality of the
one-particle potential.  This feature is distinctive from DMFT, in
which the non-locality connecting different atomic sites is often
missing, whereas  dynamical effects can be incorporated with the
one-particle potential.

While the one-shot $GW$ is a perturbative calculation starting from a
given $H_0$ (usually from the $H_0$ of Kohn-Sham Hamiltonian in
GGA/LDA), QSGW is a self-consistent perturbation method that can
determine the one-body Hamiltonian within itself.  To be clearer, let
us recall that the $GW$ approximation gives the one-particle effective
Hamiltonian whose energy dependence comes from the self-energy term
$\Sigma(\omega)$ (here we omit index of space and spin for
simplicity).  In QSGW, the static one-particle potential $\vxc$ is
generated from $\Sigma(\omega)$ as
\begin{eqnarray}
\vxc = \frac{1}{2}\sum_{ij} |\psi_i\rangle 
       \left\{ {{\rm Re}[\Sigma(\ei)]_{ij}+{\rm Re}[\Sigma(\ej)]_{ij}} \right\}
       \langle\psi_j|,
\label{eq:veff}
\end{eqnarray}
where $\ei$ and $|\psi_i\rangle$ refer to the eigenvalues and
eigenfunctions of $H_0$, respectively, and ${\rm
  Re}[\Sigma(\varepsilon)]$ is the Hermitian part of the self-energy
\cite{QSGW-PRL2004,QSGW-PRL2006,QSGW-PRB2007}.  With this $\vxc$, one
can define a new static one-body Hamiltonian $H_0$, and continue to
apply $GW$ approximation until converged.  In principle, the final
result of QSGW does not depend on the initial conditions.  Previous
QSGW studies, ranging from semiconductors
\cite{QSGW-PRL2006,QSGW-PRB2007} to the various $3d$ transition metal
oxides \cite{QSGW-PRL2006,QSGW-PRB2007,QSGW-SpinWave-JPCM2008} and
$4f$-electron systems \cite{QSGW-4f-PRB2007}, have demonstrated its
capability in the description of weakly and strongly correlated
electron materials.

\begin{figure}[t]
\begin{center}
\includegraphics[width=12cm,angle=0]{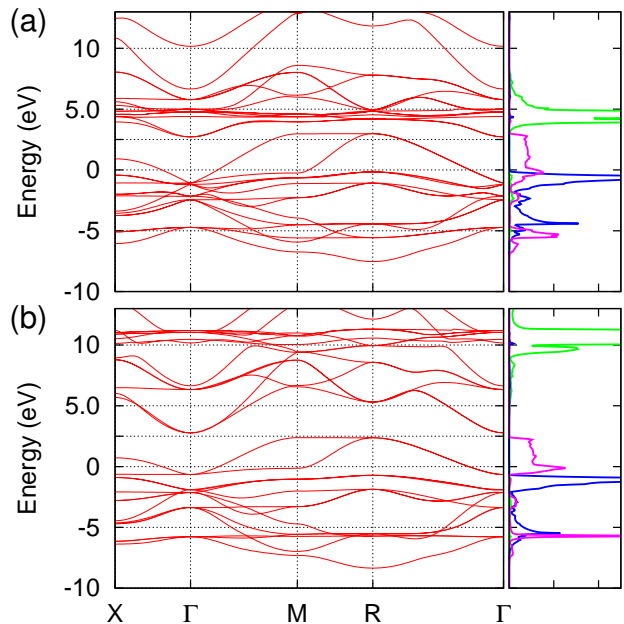}
\caption{ The calculated band dispersion of bulk LNO by (a) GGA and
  (b) QSGW. The Fermi level is set to be 0. The PDOS is also
  presented in the right panel in which the blue, magenta, and green
  lines refer to the Ni-$t_{2g}$, Ni-$e_g$, and La-$4f$ states,
  respectively. The radii of the MT spheres are 1.02 and 1.59 \AA~ for
  Ni and La, respectively.
    \label{lno-band}}
\end{center}
\end{figure}

\begin{figure}[t]
\begin{center}
\includegraphics[width=8cm,angle=0]{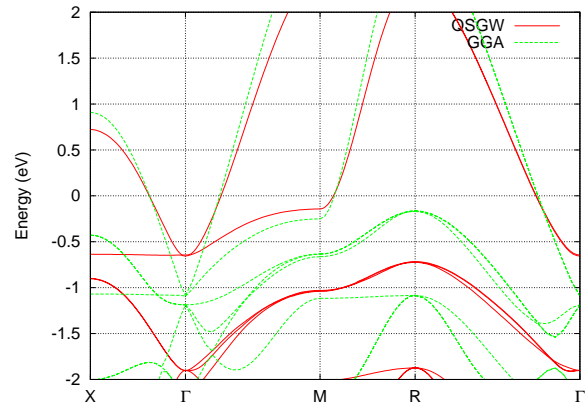}
\caption{The band dispersion of LNO near the Fermi level calculated by
  GGA (dashed green) and QSGW (solid red). The Fermi level is set to
  be 0.
  \label{LNOband_Ef}}
\end{center}
\end{figure}

\begin{table*}[t]
\begin{tabular}{c|cc|cc}
\hline\hline
System & \multicolumn{2}{|c|}{$m_{\rm QSGW}/m_{\rm GGA}$} & \multicolumn{2}{c}{$N_{\rm e_g}$} \\
\hline
        & Electron  & Hole  & GGA  & QSGW  \\
\hline
LNO     &   1.36    & 1.43  & 2.16 & 2.04 \\
LNO/LAO &   1.39    & 1.29  & 2.15 & 2.04 \\
\hline\hline
\end{tabular}
\caption{The calculated electron/hole mass ($m_{\rm QSGW}/m_{\rm
    GGA}$) and the number of $e_g$ electrons ($N_{\rm e_g}$) for LNO and
  LNO/LAO. For the effective mass of LNO, the $\Gamma$ to (R, M, X)
  and R to ($\Gamma$, M, X) lines are taken for the electron and
  hole mass, respectively, and the average values are presented. For
  LNO/LAO, $\Gamma$(M) to (X(X), M($\Gamma$)) and Z(A) to (A(R),R(Z)) 
  lines are taken for electron (hole).}\label{tab}
\end{table*}

\subsection{Computation details}
We used our new implementation of QSGW \cite{PMTQSGW-kotanikino} by
adopting the `augmented plane wave (APW) + muffin-tin orbital (MTO)',
designated by `PMT' \cite{kotani_pmt_2010,kotani_pmt_2013}, for the
one-body solver.  The accuracy of this full potential PMT method is
proven to be satisfactory in the supercell calculations of
homo-nuclear dimers from H$_2$ through Kr$_2$ with the significantly
low APW energy cutoff of $\sim$ 4 Ry by including localized MTOs
\cite{kotani_pmt_2013}. A key feature of this scheme for QSGW is that
the expansion of $\vxc$ can be made with MTOs, not APWs, which enables
us to make the real space representation of $\vxc$ at any ${\bf k}$
point. It can therefore be similar to another implementation of QSGW
based on the maximally localized Wannier functions
\cite{hamann_maximally_2009}. Also, in contrast to a previous approach
based on FP-LMTO (full potential linearized muffin-tin orbital)
\cite{QSGW-PRB2007}, our scheme is free from the fine tunings of MTO
parameters.  The basis set for the eigenfunctions can be enlarged
systematically with the APW cutoff and no empty sphere is required.

The lattice constant used for LNO (cubic perovskite) is 3.86 \AA.  For
(LNO)$_1$/(LAO)$_1$, the in-plane lattice is set to be 3.905 \AA~
(SrTiO$_3$ value as the substrate) while the $c$ lattice parameter and
the internal atomic positions were optimized with GGA.  We assumed
that the tetragonal symmetry is preserved and that rotational or
breathing-type distortion does not occur because enlarging the
unitcell requires too much computation cost for our QSGW
calculation. Since such distortions can be realized and the oxygen
cage rotations generally reduce hopping, the correlation effect in our
calculation might be underestimated as discussed further below.  We
used 8$\times$8$\times$4=256 and 8$\times$8$\times$8=512 $\bfk$ points
for the self-energy calculation in the first Brillouin zone of
(LNO)$_1$/(LAO)$_1$ and LAO, respectively. The convergency of the
present results is robust against the change of the number of {$\bf
  k$} points and the other cutoff parameters.

\section{Result and Discussion}

\subsection{Electronic structure of LaNiO$_3$}

As a prototype example that shows the limitation of the conventional
first-principle methodologies, bulk LNO has been examined by several
advanced methods. For example, Deng {\it et al.}  \cite{dmft} applied
their LDA+DMFT to bulk cubic LNO. Gou {\it et al.} performed a
comparative study using LDA, GGA, LDA$+U$ and hybrid functional
\cite{lno-hybrid}.

Our result is presented in Fig.~\ref{lno-band}. The band structure
obtained by QSGW is markedly different from GGA.  First of all, the
Ni-$e_g$ bandwidth is notably reduced. The antibonding part of the
Ni-$e_g$ dispersion reaches up to 3.0 eV in GGA result
(Fig.~\ref{lno-band}(a)) while it is 2.4 in QSGW
(Fig.~\ref{lno-band}(b)).  The bandwidth reduction is about $\sim$1.2
eV. This result clearly shows that the QSGW captures the correlation
effect which is poorly described by the conventional LDA/GGA type of
approach. Another notable difference is that two $e_g$ bands across
the Fermi level are completely decoupled from the other bands in QSGW,
which is not the case in GGA.  The bonding part is also affected. In
GGA, the bonding combination of Ni-$e_g$ is located in between $-$4.8
and $-$7.5eV (Fig.~\ref{lno-band}(a)) whereas, in QSGW, it is in
between $-$5.8 and $-$8.3 eV (Fig.~\ref{lno-band}(b)). For the bonding
part, the bandwidth change is not significant while the location is
pushed down \cite{comment1}. The $t_{2g}$ complex is also slightly
changed by QSGW.

The bandwidth reduction and the electronic correlation captured by
QSGW are important for determining material properties since the
effective mass is enhanced accordingly. It is found that the effective
electron and hole mass calculated by QSGW are about 40~\% larger than
GGA values (Table~\ref{tab}).

Notable changes are also found in the higher energy regions.  The
La-$4f$ bands located at $\sim$5 eV in GGA are pushed away by
QSGW. Since the unphysical energy position of La-$4f$ significantly
affects material properties ({\it e.g.,} bond length and binding
energy), this feature should be corrected in the LDA+$U$ or +DMFT
approach.  It causes another ambiguity in determining the additional
parameters for La-$f$ orbitals (or other rare-earth elements) besides
the transition metal $d$.  In this regard, therefore, QSGW has a clear
advantage with no adjustable parameter \cite{QSGW-LMO}. Relatively
deep core level bands (at around $-$15 eV or below; not shown) are
found to be pushed down by a few electron volts in QSGW results which
can be important for interpreting, for example, the X-ray spectroscopic
data related to such a level position.

It is instructive to compare our QSGW result with the previous DMFT by
Deng {\it et al.} \cite{dmft}. For this purpose, we present the
enlarged band dispersion around the Fermi level in
Fig.~\ref{LNOband_Ef}.  The downshift of $t_{2g}$ bands (just below
the anti-bonding $e_g$ complex), one of the main findings of
Ref.~\onlinecite{dmft}, is clearly observed in our QSGW result.
Another important feature in DMFT spectra is the `kink' at around
$-$0.2 eV which compares well with the angle-resolved photoemission
spectroscopy (ARPES) \cite{arpes}. Noticeably, QSGW somehow captures
such a feature: Compared to GGA, the reduced $e_g$ bandwidth around
the Fermi level results in the flattening of the band. In particular,
the flat band along X--$\Gamma$ at around $-$0.6 eV (see
Fig.~\ref{LNOband_Ef}) is a shade of the `kink' in DMFT and ARPES.
Specifically, we note that the DMFT QP dispersion with Im$\Sigma
=0^{+}$ goes to $-$0.6eV at the $\Gamma$ point which is in good
agreement with our QSGW dispersion. Mainly designed to construct the
optimized QP picture, QSGW does not properly include the low energy
magnetic fluctuation whereas the charge fluctuation is taken into
account within RPA \cite{phonon}. Therefore, by incorporating such an
effect in the self-energy, we expect  further flattening of the
band and the kink-like structure, being more similar with ARPES and
DMFT.

Importantly, our QSGW calculation of cubic LNO yields the
well-converged PM solution. While LDA/GGA calculations ($U$=0) predict
the PM ground state for the bulk LNO, they significantly underestimate
the correlation effects.  On the other hand, the LDA/GGA$+U$
calculation gives a magnetically ordered ground state which is in
sharp contrast to the experiment. It demonstrates a clear limitation
of the Hartree-Fock type of approach like LDA/GGA$+U$ in the
description of correlated PM phase. In this regard, QSGW has a
definitely better aspect.  Within the current implementation of QSGW,
however, it is difficult to say that the PM phase is more stable
energetically than the ferro (or anti-ferro) by comparing their total
energies. We observed that a small difference in the computational
settings can change the results regarding magnetic stability. It is
basically a numerical instability and can be remedied in principle by
improving our implementation \cite{comment-implement}. At the same
time, we presume that this instability is also related to the
intrinsic magnetic instability of the system.

\begin{figure}[t]
\begin{center}
\includegraphics[width=12cm,angle=0]{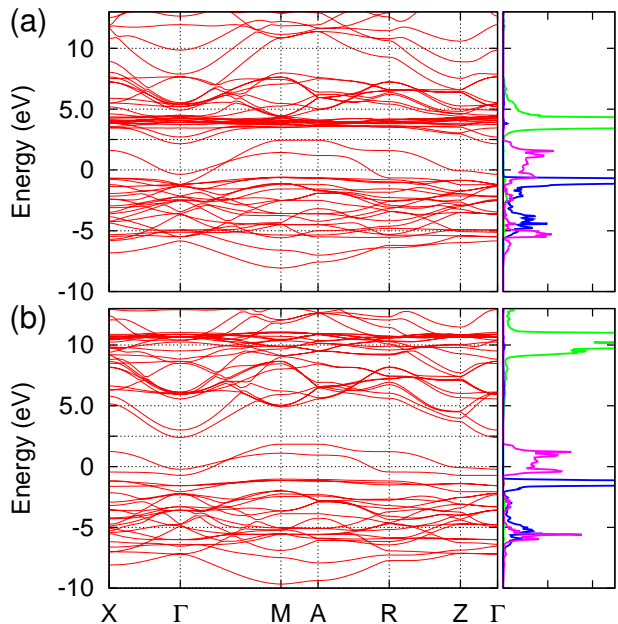} 
\end{center}
\caption{The calculated band dispersion of LNO/LAO by (a) GGA and (b)
  QSGW.  The Fermi level is set to be 0. The PDOS is also
  presented in the right panel in which the blue, magenta, and green
  lines refer to the Ni-$t_{2g}$, Ni-$e_g$, and La-$4f$ states,
  respectively.  The MT radii for Ni and La are the same with those in
  Fig.~\ref{lno-band}.
\label{lnolao-band}}
\end{figure}

\subsection{Electronic structure of (LaNiO$_3$)$_1$/(LaAlO$_3$)$_1$}

One of the most interesting issues in the LNO/LAO superlattice may be
the Fermi surface topology. According to the DMFT calculation by
Hansmann {\it et al.} \cite{Hansmann}, the inclusion of $U$ causes the
Fermi surface of LNO/LAO to be cuprate-like and only the $d_{x^2-y^2}$
band is available around the Fermi energy. However, another DMFT
calculation \cite{MJHan-DMFT} arrived at a different conclusion. Based
on the $d$--$p$ model which contains the oxygen bands explicitly in
the DMFT procedure, Han {\it et al.} obtained a different spectra;
{\it i.e.,} two $e_g$ bands across the Fermi level and the central
Fermi surface pocket is present even in the large $U$ region. Since
similarity or dissimilarity with the cuprate band structure was key to
predicting possible high-temperature superconductivity in this
superlattice, the correct description of electronic structures is of
critical importance. The different conclusions by two DMFT
calculations therefore require further examination.  Also, the
situation demonstrates the inherent ambiguity in the current DFT
(density functional theory) +DMFT formalism; {\it e.g.,} the issues of
double-counting, downfolding/projection, and the determination of $U$
and $J$.

\begin{figure}[t]
\begin{center}
\includegraphics[width=8cm,angle=0]{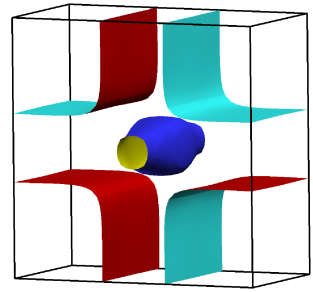}
\caption{ The Fermi surface of LNO/LAO superlattice calculated by
  QSGW, which corresponds to the band dispersion of Fig.~2(b). The
  size of central pocket is slightly increased from the GGA result
  being consistent with Ref.~\cite{MJHan-DMFT}.
    \label{FS_GW}}
\end{center}
\end{figure}

The QSGW method can be a good complementary choice as the fully charge
self-consistent method with no adjustable parameter. The calculated
band dispersion and the projected density of states (PDOS) are
presented in Fig.~\ref{lnolao-band}. The GGA band in
Fig.~\ref{lnolao-band}(a) compares well with the previous LDA or GGA
calculations \cite{MJHan-opol}. In the QSGW result presented in
Fig.~\ref{lnolao-band}(b), two Ni-$e_g$ bands around the Fermi energy
are reduced in their widths and separated from the Ni-$t_{2g}$ complex
as observed in the bulk LNO (Fig.~\ref{lno-band}(b)). The bandwidth
reduction is about 1.3 eV. In QSGW, the top of the Ni-$e_g$ bands are
clearly lower than the bottom of the upper bands with a free
electron-like feature around the $\Gamma$ point. The La-$4f$ level is
pushed above from 4 to 10.5 eV as also noted in Fig.~\ref{lno-band},
and the core level bands are pushed down by $\sim$3 eV (not shown). The
mass enhancement in the superlattice is comparable to that in LNO: The
calculated effective electron and hole mass by QSGW are $\sim$40 and
30~\% larger than the GGA values, respectively (see Table~\ref{tab}).

Most importantly, the QSGW Fermi surface has the same topology as in
GGA (see Fig.~\ref{FS_GW}).  The central Fermi surface still exists
and the topology does not support the one-band physics of
cuprates. The size of this central pocket is slightly increased from
the GGA result, which is also observed in the DMFT calculation in
Ref.~\cite{MJHan-DMFT}. This result is therefore consistent with the
DMFT result of Ref.~\onlinecite{MJHan-DMFT} rather than
Ref.~\onlinecite{Hansmann}.  Without any adjustable parameter or the
ambiguity related to down-folding or double-counting, our QSGW results
provide important new information for understanding the electronic
structure and possible high-temperature superconductivity especially
in the context of two previous DMFT studies yielding different
conclusions.

Another interesting aspect of the nickelate superlattice is the MIT
accompanied by the magnetic transition as observed by Boris {\it et
  al.} \cite{Boris-Science}. It is reported that LNO/LAO becomes
insulating and magnetic in the thin LNO limit. Since the LDA$+U$ type
of method predicts a magnetic solution even for the bulk LNO, its
predictive power is questionable in the thin LNO superlattice although
it actually gives the ferromagnetic ground state
\cite{MJHan-LDAU,Pentcheva}. On the other hand, LDA or GGA predicts
the PM phase \cite{MJHan-opol,MJHan-LDAU} while DMFT calculations are
mainly concerned with the PM region at high temperature
\cite{Hansmann,MJHan-DMFT}. In this situation QSGW can also provide
useful information. We found that the PM metallic phase is well
stabilized in the QSGW calculation while the stability of the magnetic
solution is questionable as in the case of bulk LNO (see
Sec.~III.~A). The system is likely located in the vicinity of the
magnetic instability. The calculated number of Ni-$e_g$ electrons
($N_{\rm e_g}$), which can be an important indication for
understanding MIT \cite{Nd-PRB}, is just slightly reduced by QSGW, as
summarized in Table~\ref{tab}.

\section{Discussion}
Since the important missing part in the QSGW QPs is mainly the
magnetic fluctuation and its effects are largely for the near-Fermi
energy, the overall band structure determined by QSGW should be
acceptable. At the same time, however, the following points can be
considered regarding the limitation of our calculations. First,
structural distortions (such as GdFeO$_3$-type cage rotations and
Jahn-Teller type), which were not taken into account in our
calculation to reduce the computation cost, may be critically
important. By reducing hopping, structural distortion can
significantly enhance the electronic correlation. It is also noted
that those distortions are actually found in the rare-earth
nickelates, thin films and heterostructures
\cite{MIT-RMP,Pentcheva,Chak-PRL-LNOfilm,HSKim}. Second, although we
failed to obtain a well-stabilized spin-polarized solution, it does
not necessarily mean that the magnetic solution does not exist.
Presumably the PM solution of QSGW is quite close to the magnetic
phase boundary as discussed above. Our guess is that the numerical
instability found in our spin-polarized calculation is likely related
to the intrinsic magnetic instability. Finally, the intrinsic
limitation of QSGW itself is also noted. While the charge fluctuation
is taken into account at the RPA level, the energy dependence of the
self-energy is neglected during the self-consistency cycle. Therefore,
further improvement to describe the correlation can be achieved, for
example, in combination with DMFT \cite{tomczak2012}.

\section{Summary}

QSGW calculations have been performed to understand the electronic
structure of LNO and the LNO/LAO superlattice. Without any {\it ad
  hoc} parameter or the ambiguity related to the double-counting and
downfolding issue, the effects of correlation are well described in
terms of the reduced bandwidth and the enhanced electron
mass. Importantly, the Fermi surface topology does not become
cuprate-like and one band physics is not achieved. This conclusion is
consistent with a recent DMFT calculation based on the $d$--$p$ model
and in a sharp contrast to the result from the $d$-band only
model. The PM solution especially for the superlattice is presumably
located in the vicinity of the magnetic instability, which possibly
indicates the critical role of structural distortions in the local
moment formation in this system.

\section{Acknowledgments}
We thank Prof. Hiroshi Katayama-Yoshida group at Osaka University for
hosting the helpful discussion and allowing to use the cluster
computers.  This work was supported by `Core-to-Core Program on
Computational Materias Design on Green Energy'.


\begin{thebibliography}{99}
















\bibitem{Hansmann} P. Hansmann, X. Yang, A. Toschi, G. Khaliullin,
         O. K. Andersen, and K. Held,
         Phys. Rev. Lett. {\bf 103}, 016401 (2009). 

\bibitem{MJHan-DMFT} M. J. Han, X. Wang, C. A. Marianetti, and A. J. Millis,
   Phys. Rev. Lett.  {\bf 107} 206804, (2011).

\bibitem{Boris-Science} A. V. Boris, Y. Matiks, E. Benckiser,
  A. Frano, P. Popovich, V. Hinkov, P. Wochner, M. Castro-Colin,
  E. Detemple, V. K. Malik, C. Bernhard, T. Prokscha, A. Suter,
  Z. Salman, E. Morenzoni, G. Cristiani, H.-U. Habermeier, and
  B. Keimer,
  Science {\bf 332}, 937 (2011). 

\bibitem{Hamada} N. Hamada, J. Phys. Solids  {\bf 54} 1157 (1993).

\bibitem{NNO-PRL} M. K. Stewart, J. Liu, M. Kareev, J. Chakhalian, and D. N. Basov,
    Phys. Rev. Lett.  {\bf 107} 176401, (2011).

\bibitem{Liu-PRB-RC} J. Liu, S. Okamoto, M. van Veenendaal, M. Kareev,
  B. Gray, P. Ryan, J. W. Freeland, and J. Chakhalian,
  Phys. Rev. B. {\bf 83} 161102, (2011).

\bibitem{John-epl} J. W. Freeland, J. Liu, M. Kareev, B. Gray,
         J.W. Kim, P. Ryan, R. Pentcheva, and J. Chakhalian,
  EPL {\bf 96} 57004, (2011).

\bibitem{MJHan-opol} M. J. Han, C. A. Marianetti, and A. J. Millis,
  Phys. Rev. B. {\bf 82} 134408 (2010).

\bibitem{MJHan-LDAU}  M. J. Han and M. van Veenendaal,
  Phys. Rev. B {\bf 85}, 195102 (2012).





\bibitem{QSGW-PRL2004} S. V. Faleev, M. van Schilfgaarde, and T. Kotani,
  Phys. Rev. Lett. {\bf 93}, 126406 (2004).

\bibitem{QSGW-PRL2006} M. van Schilfgaarde, T. Kotani, and S. Faleev,
  Phys. Rev. Lett. {\bf 96}, 226402 (2006).

\bibitem{QSGW-PRB2007} T. Kotani, M. van Schilfgaarde, and S. V. Faleev,
  Phys. Rev. B {\bf 76}, 165106 (2007).


\bibitem{QSGW-SpinWave-JPCM2008} T. Kotani and M. van Schilfgaarde,
  J. Phys.: Condens. Matter {\bf 20},  295214 (2008).

\bibitem{QSGW-4f-PRB2007} A. N. Chantis, M. van Schilfgaarde, and
  T. Kotani, Phys. Rev. B {\bf 76}, 165126 (2007).

\bibitem{PMTQSGW-kotanikino}
  T. Kotani and H. Kino, (unpublished).

\bibitem{kotani_pmt_2010}
  T.Kotani and Mark van Schilfgaarde, 
  Phys. Rev. B.	{\bf 81}, 125117(2010)

\bibitem{kotani_pmt_2013}
  T. Kotani and H. Kino,
  J. Phys. Soc. Jpn. {\bf 82}, 124714(2013)

\bibitem{hamann_maximally_2009}  
  D. R. Hamann, and D. Vanderbilt,
  Phys. Rev B {\bf 79}, 045109(2009)









\bibitem{dmft}  X. Deng, M. Ferrero, J. Mravlje, M. Aichhorn, and A. Georges,
  Phys. Rev. B \textbf{85},  125137 (2012).

\bibitem{lno-hybrid} G. Gou, I. Grinberg, A. M. Rappe, and J. M. Rondinelli,
  Phys. Rev. B \textbf{84}, 144101 (2011).

\bibitem{comment1} QSGW has a tendency to push down the localized bands
relative to extended bands (see Ref.~\onlinecite{QSGW-PRB2007}).

\bibitem{QSGW-LMO} T. Kotani and H. Kino, 
 J. Phys.: Condens. Matter {\bf 21}, 266002 (2009).


\bibitem{arpes} R. Eguchi, A. Chainani, M. Taguchi, M. Matsunami,
  Y. Ishida, K. Horiba, Y. Senba, H. Ohashi, and S. Shin,
  Phys. Rev. B \textbf{79}, 115122 (2009).

\bibitem{phonon} As in the case of phonon, the low magnetic
  fluctuation essentially affects the dispersion just near the Fermi
  energy.  See, for example, C. Kirkegaard, T. K. Kim, and
  Ph. Hofmann, New J. Phys. \textbf{7}, 99 (2005).

\bibitem{comment-implement} For example, we expect that the inclusion
  of magnetic fluctuations on top of the paramagnetic solution of QSGW
  can give a reasonable description.
 

\bibitem{Pentcheva} A. Blanca-Romero and R. Pentcheva,
    Phys. Rev. B {\bf 84}, 195450 (2011).





\bibitem{Nd-PRB} X. Wang, M. J. Han, L. de Medici, H. Park,
  C. A. Marianetti, and A. J. Millis, Phys. Rev. B {\bf 86}, 195136
  (2012).


\bibitem{MIT-RMP} M. Imada, A. Fujimori, and Y. Tokura,
  Rev. Mod. Phys. {\bf 70}, 1039 (1998).

\bibitem{Chak-PRL-LNOfilm} J. Chakhalian, J. M. Rondinelli, J. Liu,
  B. A. Gray, M. Kareev, E. J. Moon, N. Prasai, J. L. Cohn, M. Varela,
  I. C. Tung, M. J. Bedzyk, S. G. Altendorf, F. Strigari,
  B. Dabrowski, L. H. Tjeng, P. J. Ryan, and J. W. Freeland
  Phys. Rev. Lett. {\bf 107}, 116805 (2011).

\bibitem{HSKim} H. -S. Kim and M. J. Han, arxiv:1306.0713 (2013).


\bibitem{tomczak2012}
  J. M.Tomczak, M. van Schilfgaarde, and G. Kotliar,
  Phys. Rev. Lett. {\bf 109}, 237010 (2012)


\end{thebibliography}
\end{document}